\documentclass[aps,prd,preprintnumbers,nofootinbib,preprint]{revtex4}
\usepackage{bm}
\usepackage{latexsym}
\usepackage{dcolumn}
\usepackage{amsmath,amsfonts,amssymb}
\usepackage{graphicx,epsfig}
\usepackage{amsthm}
\usepackage{url,hyperref}

\newcommand{\be}{\begin{eqnarray}}
\newcommand{\ee}{\end{eqnarray}}
\newcommand{\bea}{\begin{eqnarray}}
\newcommand{\eea}{\end{eqnarray}}

\def\comment#1{}

\newcommand{\lp}{\ell_{\rm p}}
\newcommand{\mpl}{m_{\rm p}}

\usepackage{color}

\definecolor{darkred}{rgb}{.8,0,0}

\definecolor{darkblue}{rgb}{0,0,.7}

\definecolor{darkgreen}{rgb}{0,.7,0}

\begin{document}

%
%
%%%%%%%%%%%%%%%%%%%%%%%%%%%%%%%%%%%%%%%%%%%%%%%%%%%%%%%%%%%%%%
\title{GUP and the  no-cloning theorem} 
%%%%%%%%%%%%%%%%%%%%%%%%%%%%%%%%%%%%%%%%%%%%%%%%%%%%%%%%%%%%%%
%
%
%
%
%
\author{{Elias~C.~Vagenas}$^1$}\email[email:~]{elias.vagenas@ku.edu.kw}
\author{Ahmed~Farag~Ali$^{2,3}$} \email[email:~]{ahmed@quantumgravityresearch.org}
\author{Hassan~Alshal$^{4,5}$\vspace{1ex}}\email[email:~]{halshal@sci.cu.edu.eg}
\affiliation{$^1$Theoretical Physics Group, Department of Physics, Kuwait University, P.O. Box 5969, Safat 13060, Kuwait\vspace{1ex}}
\affiliation{$^2$Department of Physics, Faculty of Science, Benha University, Benha, 13518, Egypt\vspace{1ex}}
\affiliation{$^3$Quantum Gravity Research, Los Angeles, CA 90290, USA \vspace{1ex}}
\affiliation{$^4$Department  of Physics, Faculty of Science, Cairo University, Giza, 12613, Egypt
\vspace{1ex}}
\affiliation{$^5$Department of Physics, University of Miami, Coral Gables, FL 33146, USA}
%
%
%
%
%\date{\today}
%
%%%%%%%%%%%%
%%%%%%%%%%%%
\begin{abstract}
%%%%%%%%%%%%
%
%
%
%
%
%
\par\noindent
Motivated by a recent work by Yongwan Gim, Hwajin Um, and Wontae Kim, we investigate the validity of the no-cloning theorem in the context of generalized uncertainty principle. In particular, in the presence of linear and quadratic terms of momentum in generalized uncertainty principle, we first compute the energy density at a given modified temperature and then using the modified Stefan-Boltzmann law we derive the modified Page time. Finally, we calculate the modified required energy for the information to be encoded into a message and be sent to an observer inside the black hole.\\
\end{abstract}
%
%\pacs{03.65.Ta, 03.65-w}
%
%
%
\maketitle
%
%
%
%
%
%
%%%%%%%%%%%%%%%
\section{Introduction}
%%%%%%%%%%%%%%%
%
%
%
%
\par\noindent
In 1974, S.W. Hawking decided to introduce black holes to the quantum world \cite{Hawking:1974rv}. This ``marriage" of General Relativity (GR) and Quantum Mechanics (QM) proved that black holes are not really black but they radiate (Hawking radiation) \cite{Hawking:1974sw}. At the same time, the Information Loss Paradox was revealed \cite{Hawking:1976ra} and though several resolutions have been proposed \cite{Giddings:2007pj, Mathur:2009hf, 
Almheiri:2012rt, Papadodimas:2012aq}, it still remains  a yet-to-be-solved problem.
\par\noindent
One of the resolutions is based on Hawking radiation and in particular, one assumes that the Hawking radiation encodes the black hole information. In this context, an infalling observer (Alice) is crossing the event horizon of the black hole carrying all information of the infalling matter state. Outside of the horizon, the local observer (Bob)  collects all the information of  the infalling matter state which is reflected on the horizon in the form Hawking radiation. If after time at least equal to 
Page time \cite{Page:1993wv}, Bob decides to dive in the black hole then Alice can send him  all the information encoded in a message. Thus, Bob will have in his hands two copies of the information which will be a violation of the no-cloning theorem \cite{Wootters:1982zz}. 
The no-cloning theorem  was saved by Susskind, Thorlacius and Uglum \cite{Susskind:1993if} who proposed that there is a stretched horizon just outside of the event horizon. The information of the infalling matter state heats up the stretched horizon which reradiates the information as Hawking radiation.  Therefore, a copy of information is already inside the black hole heading to the singularity and one more copy is radiated outside the horizon. However, there is no way an observer to observe both copies simultaneously time since this will violate the principle of complementarity which is one of the principles of QM. In particular, if one perform a gedanken experiment in  the Schwarzschild black hole  spacetime the required energy to encode in a  message all information that Alice carries  and send this message to Bob when he is inside the black hole has to be super-Planckian \cite{Susskind:1993mu}. 
\par\noindent
Of course, one may claim that  the marriage between GR and QM has issues because we are trying to keep all principles of QM untouched. Maybe it is necessary to modify QM and not GR. In this line of thought, one considers modifications of the Heisenberg Uncertainty Principle (HUP) known as Generalized Uncertainty Principle (GUP) \cite{Veneziano:1986zf, Gross:1987ar, Amati:1988tn, Konishi:1989wk, Maggiore:1993rv, Garay:1994en, Scardigli:1999jh, Adler:2001vs, Das:2008kaa, Ali:2009zq}. So, one may suggest that GUP can play a crucial role in the solution of the Information Loss Paradox \cite{Itzhaki:1995tc, Chen:2014bva, Gim:2017rmn} and this work aims to shed some light on this direction.
\par
The remainder of this paper is organized as follows. In Section II, we briefly derive the expressions for the Page time and the required energy in order  the information Alice carries to be encoded into a message and to be sent to Bob inside the Schwarzschild black hole in the framework of HUP and quadratic GUP. In Section III, we calculate the Page time and the required energy in the presence of linear and quadratic terms in momentum of GUP. Finally, in Section IV we briefly present our results and  some concluding comments are given. In this work, we use the natural units $\hbar=c=1$.
%
%
%
%
%
%
%%%%%%%%%%%%%%%%%%%%%%%%%%%%%%%
\section{HUP, quadratic GUP, and the non-cloning theorem}
%%%%%%%%%%%%%%%%%%%%%%%%%%%%%%%
%
%
%
\par\noindent
In this section, we will briefly summarize the derivation of Page time, i.e., $t_{p}$, and that of the required energy, i.e., $\Delta E$, as presented in Ref. \cite{Gim:2017rmn} . This energy is needed in order the information that Alice carries to be encoded into a message and then to be sent to Bob who is already inside the Schwarzschild black hole. For this reason, we utilize the Kruskal-Szekeres diagram of the Schwarzschild black hole \cite{Townsend:1997ku}. 
The Schwarzschild black hole metric in Kruskal-Szekeres coordinates reads
\be
ds^2 = -\frac{32G^{3}M^{3}}{r}e^{-\frac{r}{2GM}}dUdV
\label{metric}
\ee
with $M$ to be the black hole mass, $U=\pm e^{-\frac{(t-r^{*})}{4GM}}$ and $V= e^{\frac{(t+r^{*})}{4GM}}$ are 
the Kruskal-Szekeres coordinates and $r^{*}=r+2GM \ln\left\vert \frac{r-2GM}{2GM}\right\vert$  is the Regge-Wheeler radial coordinate. It is well known that the area, i.e., $A$, and Hawking temperature, i.e., $T_{H}$, of the Schwarzschild black hole are given as
\bea
A&=&16\pi G^2 M^2\\
T_{H}&=& \frac{1}{8\pi GM}
\label{temp1}
\eea
while the Stefan-Boltzman law reads 
\be
\frac{dM}{dt}=-\sigma A T^{4}
\label{stefan1}
\ee
\par\noindent
where $\sigma$ is the Stefan-Boltzman constant.
\par\noindent
As already mentioned in the Introduction, Alice will travel from the exterior quadrant ($\mathbf{I}$) into a Schwarzschild black hole singularity through the black hole interior quadrant ($\mathbf{II}$). She will enter the horizon at  $V_A$ and will practice free fall inside the black hole after the horizon but still far away from singularity, i.e., she is in an inertial frame according to GR. Using a Hawking radiation detector, Bob is communicating with Alice while she is getting closer to the stretched horizon.  Bob is standing just outside the stretched horizon not doing anything and will enter the horizon at $V_B$. In a time that is greater or equal to Page time, Alice has to send her messages to Bob before he meets the singularity so $U_{A}=U_{B}=V^{-1}_{B}=e^{-t_{p}/4GM}$. Using the metric given in Eq. (\ref{metric}), one can compute the proper time $\Delta\tau$ that Alice needs to send messages to Bob who remains near the horizon at $U_A$ and the horizon is located at $r_{H}=2GM$ . Thus, we consider that $\Delta U_{A} = U_A $ and $\Delta V_{A}$ near the horizon gets a fixed value.\\
%
%
%
%%%%%%%%%%
\par\noindent
{\bf i.  HUP}
%%%%%%%%%%
%
%
%
%
\par\noindent
In this case, there are no quantum-gravity corrections, so the HUP is of the form
\be
\Delta x \Delta p \geq 1
\label{gup1}
\ee
and in terms of energy and time becomes
\be
\Delta E \Delta \tau \geq 1~.
\ee
\par\noindent
Employing the Stefan-Boltzmann law and Hawking temperature, the Page time reads
\be
t_{p} \sim G^{2} M^{3}
\label{pagetime1}
\ee
and the proper time will be
\be
\Delta \tau^{2} \sim G^{2} M^{2} e^{-G M^{2}}~.
\label{proper1}
\ee
Therefore, the required energy for Alice to encode all information in a message and send it to Bob will be
\be
\Delta E \sim \frac{1}{GM} e^{GM^2}~. 
\label{energy1}
\ee
\par\noindent
It is evident that the required energy is larger than the black hole mass, i.e., $\Delta E \gg M$, so Alice needs super-Planckian energies to encode her message and send it to Bob. Therefore, there is no violation of the no-cloning theorem.\\
%
%
%
%%%%%%%%%%%%
\par\noindent
{\bf ii.  Quadratic GUP}
%%%%%%%%%%%%
%
%
\par\noindent
In this case, we consider the GUP with a quadratic term in momentum \cite{Veneziano:1986zf, Gross:1987ar, Amati:1988tn, Konishi:1989wk, Maggiore:1993rv, Garay:1994en, Scardigli:1999jh}, namely 
\be
\Delta x \Delta p  \geq 1 + \alpha_{GUP}  \lp^{2}  \Delta p^{2} 
\label{gup2}
\ee
with $\alpha_{GUP}$ to be the dimensionless GUP parameter and the Planck length to be $\lp = \sqrt{G}$. 
It is obvious that the temperature of the black hole will no longer be the Hawking temperature but a modified 
one \cite{Adler:2001vs} which is of the form 
\be
T=\frac{1}{8\pi GM}+ \frac{\alpha_{GUP}}{32\pi G^{2} M^{3}}~.
\label{temp2}
\ee
\par\noindent
Employing the Stefan-Boltzmann law and Hawking temperature, the Page time reads \cite{Gim:2017rmn}
\be
t_{p} \sim G^{2} M^{3} - \alpha_{GUP} G M
\label{pagetime2}
\ee
and the proper time will now be computed from the metric given by Eq. (\ref{metric}), if Eq. (\ref{pagetime2}) 
is substituted in 
\be
\Delta \tau^{2} \sim G^{2} M^{2} e^{-\frac{t_{p}}{GM}}~.
\label{proper2}
\ee
Therefore, the required energy for Alice to encode all information in a message and send it to Bob will be 
\cite{Gim:2017rmn}
\be
\Delta E \sim \frac{M}{2\alpha_{GUP}} e^{-GM^{2}+\alpha_{GUP}} \left[ 1-\sqrt{1-\frac{4\alpha_{GUP}}{GM^{2}}
 e^{2\left( GM^{2}-\alpha_{GUP} \right)}}\right] ~. 
\label{energy2}
\ee
\par\noindent
It can be shown graphically that the required energy is larger than the black hole mass, i.e., $\Delta E \gg M$, so Alice needs super-Planckian energies to encode her message and send it to Bob. Therefore, there is no violation of the no-cloning theorem.\\
%
%
%
%
%
%
%
%
%
%
%%%%%%%%%%%%%%%%%%%%%%%%%%%%%%%%%
\section{Linear and quadratic GUP and the no-cloning theorem}
%%%%%%%%%%%%%%%%%%%%%%%%%%%%%%%%%
%
%
%
\par\noindent
Following the analysis in Ref. \cite{Gim:2017rmn}, we will now calculate the corrected required energy for information cloning, when a version of GUP with linear and \textit{quadratic} terms \cite{Ali:2009zq} is utilized. The linear and quadratic GUP is of the form
\be
\Delta x \Delta p \geqslant \frac{1}{2} \left[1 + \left( \frac{\alpha}{\sqrt{\left\langle p^2 \right\rangle }} 
+ 4 \alpha^2 \right) (\Delta  p)^2 + 4 \alpha^2 \left\langle p \right\rangle^2 - 2 \alpha\sqrt{\left\langle p^2 \right\rangle } \, \right]
\label{gup3}
\ee
\par\noindent
where $\alpha = \alpha_0 \lp = \alpha_o / \mpl$ with $\alpha_0$ to be the dimensionless GUP constant 
\footnote{To avoid confusion upon comparison of our work with Ref.   \cite{Gim:2017rmn}, our dimensionless GUP parameter $\alpha_0$ is related to the corresponding $\alpha_{GUP}$ of Ref.   \cite{Gim:2017rmn} through the equation $ \alpha_{0}^2 = \alpha_{GUP}$ .} and  
$\ell_p=m_p^{-1}= \sqrt{G}$. 
\par\noindent
For symmetry purposes, we are interested in mirror-symmetric states, i.e., $\left\langle p  \right\rangle ^{2} =0$, hence  $(\Delta p)^2 = \left\langle p^2 \right\rangle $, thus
\be
\Delta x \Delta p  \simeq \left[ 1  -  \alpha  \Delta p  + 4 \alpha^{2} (\Delta  p)^{2}   \right]
\label{gup4}
\ee
\par\noindent
which renders momentum corresponding to the fundamental length as
\be 
\Delta p = \left[ \frac{\Delta x + \alpha}{8 \alpha^2} \right] \left[ 1 \pm \sqrt{1 - 
\frac{16 \alpha^2}{\left( \Delta x + \alpha \right)^2}} \, \right].
\label{gup4mom}
\ee
\par\noindent
The Hawking radiation emitted from a Schwarzschild black hole with mass $M$ and event horizon radius at $r_{Sch} = 2MG$ will be considered as photons  bounded in a cube of length $L$. These photons will satisfy a GUP-modified de Broglie relation which in the light of Eq. (\ref{gup4}), their average wavelength $\lambda$ set equal to the minimum uncertainty on position, i.e., $\Delta x_{min}$, now reads
\be
\lambda \simeq \frac{1}{p} (1 - \alpha p + 4 \alpha^2 p^2)~.
\label{lambda}
\ee
\par\noindent 
At this point, it should be noted that in Ref. \cite{Gim:2017rmn} there is no linear term in momentum, i.e., $(-\alpha p)$, hence it is expected that this term will affect the rest of calculation of the energy. The corresponding frequency  becomes
\be
\nu \simeq \lambda^{-1} \simeq p \ (1 + \alpha p - 3 \alpha^2 p^2 + \mathcal{O} (\alpha^3))~.
\label{frequency1}
\ee
\par\noindent
It is obvious that for such de Broglie waves ($E = p$) the frequency can be written as
\be
\nu \simeq E \ (1 + \alpha E - 3 \alpha^2 E^2 + \mathcal{O} (\alpha^3))~.
\label{frequency2}
\ee
\par\noindent
At this point, it is crucial to recognize the constraint $\displaystyle{0\leq E\leq\frac{1}{6\alpha}[1+\sqrt{13}]}$ imposed 
by Eq.  (\ref{frequency2})  in order to avoid  the negative frequencies. 
\par\noindent
To compute the Page time, one has first to compute the energy density of the oscillating photon modes in the cube. 
Therefore, we will follow a similar analysis with the one for the derivation of Stefan-Boltzmann law. Specifically, the standard energy density is defined as \cite{BB}
\bea
\rho &=& \frac{1}{V}\int \bar{E}\, g(\nu)d\nu \\
&=& 2\int \bar{E}\, d^{3}\nu
\label{density1}
\eea
where $g(\nu)$ is the number of oscillating photon modes in the infinitesimal interval $[\nu, \nu+d\nu]$ and 
$\bar{E}$ is the average energy per oscillating photon mode which reads
\be\label{E}
\bar{E}=\frac{E}{e^{\frac{E}{T}}-1}~.
\ee
\par\noindent
In addition, from  Eq. (\ref{frequency2}), it is easily seen that 
\be\label{v}
d^3 \nu = 4 \pi \nu^2 d \nu \simeq 4\pi E^2 (1 + 4 \alpha E - 10 \alpha^2 E^2 + \mathcal{O} (\alpha^3)) \ dE~.
\label{frequency3}
\ee
\par\noindent
Substituting Eq. (\ref{E}) and Eq. (\ref{v}) into Eq.(\ref{density1}), the energy density of such ``Planckian" photons reads
\be
\rho \simeq 8\pi \int dE \ \ \frac{E^3(1 + 4\alpha E - 10 \alpha^2 E^2)}{[e^ {\frac{E}{T} }- 1]}~.
\label{density2}
\ee
\par\noindent
As a reminder, Eq. (\ref{density2}) is \textit{bona fide} up to energy cutoff $\displaystyle{E\leq E_{max} = \frac{1}{6\alpha}[1+\sqrt{13}]}$. Then upon change of variable $\xi = E/T$, energy density now becomes
\bea\label{bose-integrals}
\rho (T) & \simeq & 8\pi \left[  T^4 \int d \xi \frac{\xi^3}{e^{\xi} - 1} + 4 \alpha T^5 \int d \xi \frac{\xi^4}{e^{\xi} - 1} - 
10 \alpha^2 T^6 \int d \xi \frac{\xi^5}{e^{\xi} - 1}\right] \\
& \simeq& 8\pi \left[ \frac{\pi^4}{15}T^{4 } + 4\alpha \left(\frac{24\pi^5}{295}\right) T^{5} - 10\alpha^2
 \left(\frac{8\pi^6}{63}\right)T^{6} \right]
 \label{density3}
\eea
with the temperature $T$ to be the GUP-modified one since we are working now 
in the framework of linear and quadratic GUP. 
 It is noteworthy that the three integrals that appear in the RHS of Eq. (\ref{bose-integrals}) are all Bose integrals. The Bose integral, $I_{B}(n)$, is defined as
\be
I_{B}(n) = \int^{\infty}_{0}dx\, \frac{x^{n}}{e^{x}-1}
\ee
which is equal to
\be
I_{B}(n) =  \zeta(n+1)\,\Gamma(n+1)
\ee
with $\zeta(n)$ to be the Riemann zeta function and $\Gamma(n)$ is the gamma function.
\par\noindent
The specific expression for the GUP-modified temperature is given in Ref.  \cite{Vagenas:2018zoz}  by
\bea
T & \simeq & \frac{1}{GM} \left[ 1 - \frac{\alpha}{GM} + \frac{\alpha^2}{G^2 M^2} \right] \\
& \simeq & \frac{1}{GM} \left[ 1 - \frac{\alpha}{GM} \left( 1 - \frac{\alpha}{GM} \right) \right].
\label{temp3}
\eea
\par\noindent
Furthermore, the GUP-modified Stefan-Boltzmann law now reads
\be
 \frac{dM}{dt} \simeq -A \rho(T) \simeq (GM)^2  \rho(T) 
 \label{stefan2}
 \ee
 and employing Eq. (\ref{density3})  and Eq. (\ref{temp3}), Page time for the Schwarzschild black hole now becomes
\bea 
t_{p}(M)   & \simeq& \int dM \frac{(GM)^{-2}}{ \rho(T)} \\
& \simeq  & \int dM \frac{1}{G^2M^2 \left(T^4 + \alpha T^5 - \alpha^2 T^6 \right) }~. \label{eq12}\\
& \simeq  &  \frac{1}{3}  G^{2} M^{3} +  \frac{3}{2}\alpha G M^2 + 5 \alpha^2 M  ~.
\label{pagetime3}
\eea
\par\noindent
It is noteworthy that here we have assumed that the uncertainty in the position of photons when in the black hole, is 
proportional to the size of the black hole and thus the photon wavelength is $\lambda \simeq \Delta x = r_{H}=2GM$.
\par\noindent
Utilizing Eq. (\ref{proper2}) and the Page time given by Eq. (\ref{pagetime3}), the proper time now reads 
\be
\Delta \tau \simeq  GM \exp \left[-\frac{1}{3}  G M^{2} -  \frac{3}{2}\alpha_{0} G^{1/2}  M - 5 \alpha_{0}^2   \right] ~.
\label{proper3}
\ee
%
%
%
%
%
%
%%%%%%%%%%%%%%%%%%%%%%%%%%%%%%%%%%%%%
\begin{figure}

\includegraphics[width=15cm, height=10cm]{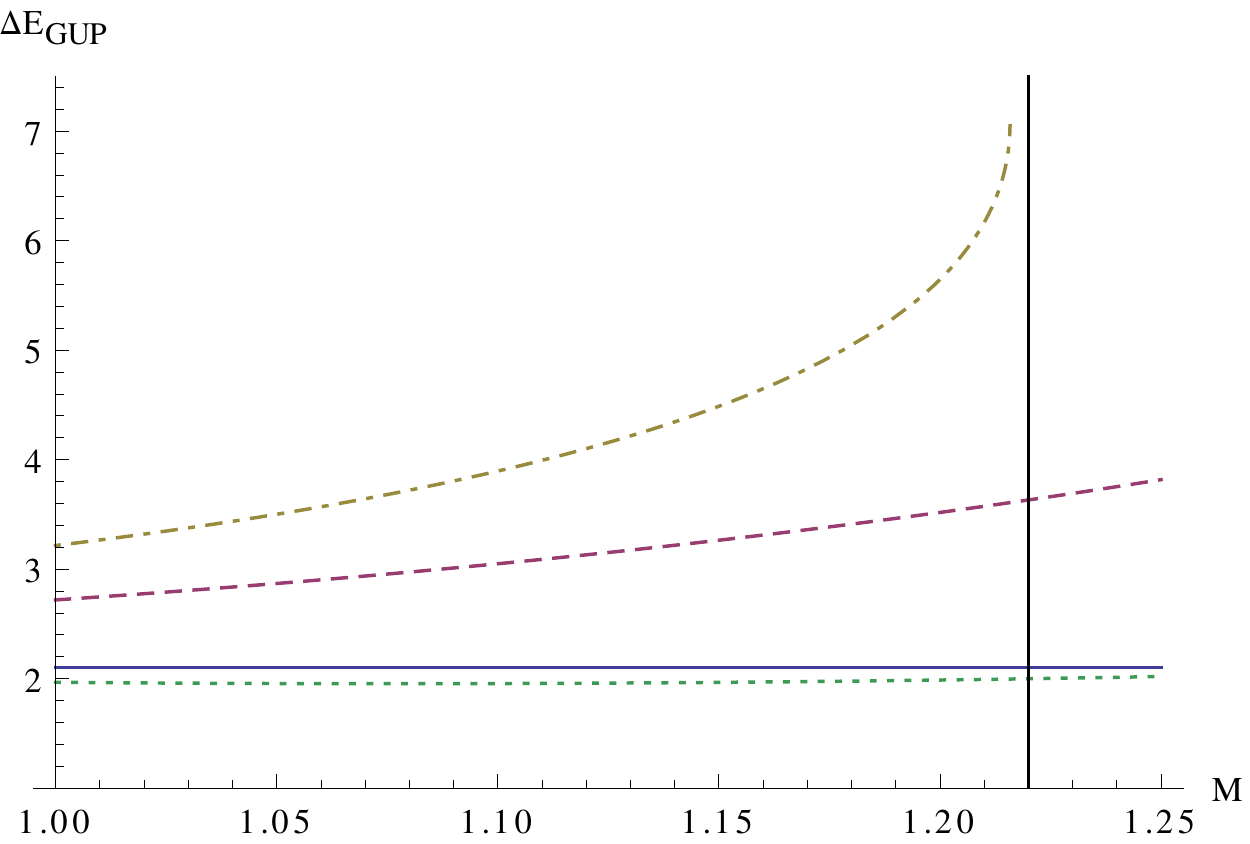}\\

\caption{In this figure,  we have used Planck units, i.e.,  $G=c=\hbar=1$, so $\ell_p=m_p^{-1}= 1$. The solid line describes the case in which the energy $\Delta E$ equals the black hole mass, i.e., M.  The dashed line depicts the required energy when we consider the HUP, i.e. $\alpha=0$. The dotted dashed line depicts the required energy when the quadratic GUP is considered. Finally, the dotted line below the solid line provides the required energy $\Delta E$ when the linear as well as the quadratic terms in momentum are present in the GUP. The GUP parameter has been set equal to $\alpha=\sqrt{0.02}$.}
\end{figure}
%%%%%%%%%%%%%%%%%%%%%%%%%%%%%%%%%%%%%%
%
%
%
%
%
\par\noindent
The required energy  for Alice to encode information in a message and send it to Bob is now
\be 
\Delta E =  \frac{\Delta \tau + \alpha_{0} G^{1/2}}{8 \alpha_{0}^{2} G} \times   
\left[  1 - \sqrt{1 - \left[ \frac{4 \alpha_{0} G^{1/2}}{ \Delta \tau + \alpha_{0} G^{1/2}} \right]^2} \right ]
\label{energy3}
\ee
\par\noindent
and by expanding in terms of the dimensionless GUP parameter $\alpha_{0}$, the proper time 
now becomes
\be
\Delta E \simeq \frac{1}{\Delta \tau} \left(1-\alpha_{0} \frac{G^{1/2}}{\Delta \tau}+ 5 \alpha_{0}^{2} \frac{G}{\Delta \tau^2} \right).
\label{energy4}
\ee
\par\noindent
It is evident that one cannot decide if this energy required to send the message from Alice to Bob  is more or less than the black hole mass when the linear and quadratic GUP terms are taken into account. However, using Fig. 1 one can graphically show that the required energy is less than the black hole mass, i.e., $\Delta E < M$. Therefore, the duplication of information is possible and the no-cloning theorem is violated while the black hole complementarity has to be restated or to be abandoned as a resolution to Information Loss Paradox.
%
%
%
%
%
%
%
%
%
%
%
%
%%%%%%%%%%%%%%%%%%%%
\section{Conclusions}
%%%%%%%%%%%%%%%%%%%%
%
%
%
%
\par\noindent
In this work, we were motivated by the work by Yongwan Gim, Hwajin Um, and Wontae Kim \cite{Gim:2017rmn}, 
and following their analysis we found the Page time as well as the required energy in order the infalling observer named Alice to encode all information she carries into a message and send it to a local observer named Bob when he is inside the black hole. Bob was initially standing was just outside the stretched horizon and was collecting the information Alice carries via the Hawking radiation which stems from the stretched horizon which is heated up by the information of the infalling matter state. We have shown that in the context of GUP with linear and quadratic terms in momenta, Bob can simultaneously hold in his hands the information from Alice inside the black hole and the same information which was collected by him while standing outside the stretched horizon. Thus, adopting the syllogism developed in Ref. \cite{Gim:2017rmn} the duplication of information is possible.
%and the no-cloning theorem is violated. 
\par\noindent
Therefore, the statement that the no-cloning theorem is safe even in the presence of GUP is inaccurate. Our present paper stands as a counter-example to this statement.

Furthermore, it seems that the black hole complementarity which was proposed as a resolution for the Information Loss Paradox  has to be restated or to be completely abandoned. However, we believe that  the Information Loss Paradox which is related to such a fundamental concept such as the information of a system can not be resolved using semiclassical methods or phenomenological models 
of Quantum Gravity. The resolution to such foundational problems can be offered only by a concrete theory of Quantum Gravity which is still to be developed.
%
%
%
%
%%%%%%%%%%%%%%%%%%%%%%%%%%%
\section{Acknowledgements}
\par\noindent
The authors  would like to thank Wontae Kim and Yongwan Gim for their useful feedback.
%%%%%%%%%%%%%%%%%%%%%%%%%%%
%
%
%
%
%
%
%
%
%%%%%%%%%%%%%%%
%
%
%
%
%
%
%
%
%%%%%%%%%%%%%%%%%

%%%%%%%%%%%%
%
%
%%%%%%%%%%%%%%%%%%%

\begin{thebibliography}{99}
%%%%%%%%%%%%%%%%
%
%
%
%\cite{Hawking:1974rv}
\bibitem{Hawking:1974rv} 
  S.~W.~Hawking,
  %``Black hole explosions,''
  Nature {\bf 248}, 30 (1974).


%\cite{Hawking:1974sw}
\bibitem{Hawking:1974sw} 
  S.~W.~Hawking,
  %``Particle Creation by Black Holes,''
  Commun.\ Math.\ Phys.\  {\bf 43}, 199 (1975)
  Erratum: [Commun.\ Math.\ Phys.\  {\bf 46}, 206 (1976)].


%\cite{Hawking:1976ra}
\bibitem{Hawking:1976ra} 
  S.~W.~Hawking,
  %``Breakdown of Predictability in Gravitational Collapse,''
  Phys.\ Rev.\ D {\bf 14}, 2460 (1976).
  
  
 %\cite{Giddings:2007pj}
\bibitem{Giddings:2007pj} 
  S.~B.~Giddings,
  %``Black holes, information, and locality,''
  Mod.\ Phys.\ Lett.\ A {\bf 22}, 2949 (2007)
  %doi:10.1142/S0217732307025923
  [arXiv:0705.2197 [hep-th]].
  
  
    %\cite{Mathur:2009hf}
\bibitem{Mathur:2009hf} 
  S.~D.~Mathur,
  %``The Information paradox: A Pedagogical introduction,''
  Class.\ Quant.\ Grav.\  {\bf 26}, 224001 (2009)
  %doi:10.1088/0264-9381/26/22/224001
  [arXiv:0909.1038 [hep-th]].
  
  
  %\cite{Almheiri:2012rt}
\bibitem{Almheiri:2012rt} 
  A.~Almheiri, D.~Marolf, J.~Polchinski and J.~Sully,
  %``Black Holes: Complementarity or Firewalls?,''
  JHEP {\bf 1302}, 062 (2013)
  %doi:10.1007/JHEP02(2013)062
  [arXiv:1207.3123 [hep-th]].
 
  
  
  %\cite{Papadodimas:2012aq}
\bibitem{Papadodimas:2012aq} 
  K.~Papadodimas and S.~Raju,
  %``An Infalling Observer in AdS/CFT,''
  JHEP {\bf 1310}, 212 (2013)
 % doi:10.1007/JHEP10(2013)212
  [arXiv:1211.6767 [hep-th]].



%\cite{Page:1993wv}
\bibitem{Page:1993wv} 
  D.~N.~Page,
  %``Information in black hole radiation,''
  Phys.\ Rev.\ Lett.\  {\bf 71}, 3743 (1993)
  %doi:10.1103/PhysRevLett.71.3743
  [hep-th/9306083].


%\cite{Wootters:1982zz}
\bibitem{Wootters:1982zz} 
  W.~K.~Wootters and W.~H.~Zurek,
  %``A single quantum cannot be cloned,''
  Nature {\bf 299}, 802 (1982).
  
  

%\cite{Susskind:1993if}
\bibitem{Susskind:1993if} 
  L.~Susskind, L.~Thorlacius and J.~Uglum,
  %``The Stretched horizon and black hole complementarity,''
  Phys.\ Rev.\ D {\bf 48}, 3743 (1993)
  %doi:10.1103/PhysRevD.48.3743
  [hep-th/9306069].


%\cite{Susskind:1993mu}
\bibitem{Susskind:1993mu} 
  L.~Susskind and L.~Thorlacius,
  %``Gedanken experiments involving black holes,''
  Phys.\ Rev.\ D {\bf 49}, 966 (1994)
  %doi:10.1103/PhysRevD.49.966
  [hep-th/9308100].



%\cite{Veneziano:1986zf} 
\bibitem{Veneziano:1986zf} 
  G.~Veneziano,
  %``A Stringy Nature Needs Just Two Constants,''
  Europhys.\ Lett.\  {\bf 2}, 199 (1986).
  %doi:10.1209/0295-5075/2/3/006
  %%CITATION = doi:10.1209/0295-5075/2/3/006;%%
  %348 citations counted in INSPIRE as of 19 Jan 2016


%\cite{Gross:1987ar}
\bibitem{Gross:1987ar} 
  D.~J.~Gross and P.~F.~Mende,
  %``String Theory Beyond the Planck Scale,''
  Nucl.\ Phys.\ B {\bf 303}, 407 (1988).
  %doi:10.1016/0550-3213(88)90390-2
  %%CITATION = doi:10.1016/0550-3213(88)90390-2;%%
  %810 citations counted in INSPIRE as of 19 janv. 2016



%\cite{Amati:1988tn}
\bibitem{Amati:1988tn} 
  D.~Amati, M.~Ciafaloni and G.~Veneziano,
  %``Can Space-Time Be Probed Below the String Size?,''
  Phys.\ Lett.\ B {\bf 216}, 41 (1989).
  %doi:10.1016/0370-2693(89)91366-X
  %%CITATION = doi:10.1016/0370-2693(89)91366-X;%%
  %741 citations counted in INSPIRE as of 20 Nov 2015
    

%\cite{Konishi:1989wk}
\bibitem{Konishi:1989wk} 
  K.~Konishi, G.~Paffuti and P.~Provero,
  %``Minimum Physical Length and the Generalized Uncertainty Principle in String Theory,''
  Phys.\ Lett.\ B {\bf 234}, 276 (1990).
  %doi:10.1016/0370-2693(90)91927-4
  %%CITATION = doi:10.1016/0370-2693(90)91927-4;%%
  %315 citations counted in INSPIRE as of 19 janv. 2016




%\cite{Maggiore:1993rv}
\bibitem{Maggiore:1993rv} 
  M.~Maggiore,
  %``A Generalized uncertainty principle in quantum gravity,''
  Phys.\ Lett.\ B {\bf 304}, 65 (1993)
  %doi:10.1016/0370-2693(93)91401-8
  [hep-th/9301067].
  %%CITATION = doi:10.1016/0370-2693(93)91401-8;%%
  %427 citations counted in INSPIRE as of 20 Nov 2015
    



   
%\cite{Garay:1994en}
\bibitem{Garay:1994en} 
  L.~J.~Garay,
  %``Quantum gravity and minimum length,''
  Int.\ J.\ Mod.\ Phys.\ A {\bf 10}, 145 (1995)
  %doi:10.1142/S0217751X95000085
  [gr-qc/9403008].
  %%CITATION = doi:10.1142/S0217751X95000085;%%
  %740 citations counted in INSPIRE as of 20 Nov 2015


%\cite{Scardigli:1999jh}
\bibitem{Scardigli:1999jh} 
  F.~Scardigli,
  %``Generalized uncertainty principle in quantum gravity from micro - black hole Gedanken experiment,''
  Phys.\ Lett.\ B {\bf 452}, 39 (1999)
  %doi:10.1016/S0370-2693(99)00167-7
  [hep-th/9904025].
  %%CITATION = doi:10.1016/S0370-2693(99)00167-7;%%
  %302 citations counted in INSPIRE as of 20 Nov 2015
   
   
   
 %\cite{Adler:2001vs}
\bibitem{Adler:2001vs} 
  R.~J.~Adler, P.~Chen and D.~I.~Santiago,
  %``The Generalized uncertainty principle and black hole remnants,''
  Gen.\ Rel.\ Grav.\  {\bf 33}, 2101 (2001)
 % doi:10.1023/A:1015281430411
  [gr-qc/0106080].


%\cite{Das:2008kaa}
\bibitem{Das:2008kaa} 
  S.~Das and E.~C.~Vagenas,
  %``Universality of Quantum Gravity Corrections,''
  Phys.\ Rev.\ Lett.\  {\bf 101}, 221301 (2008)
 % doi:10.1103/PhysRevLett.101.221301
  [arXiv:0810.5333 [hep-th]].



%\cite{Ali:2009zq}
\bibitem{Ali:2009zq} 
  A.~F.~Ali, S.~Das and E.~C.~Vagenas,
  %``Discreteness of Space from the Generalized Uncertainty Principle,''
  Phys.\ Lett.\ B {\bf 678}, 497 (2009)
  %doi:10.1016/j.physletb.2009.06.061
  [arXiv:0906.5396 [hep-th]].




%\cite{Itzhaki:1995tc}
\bibitem{Itzhaki:1995tc} 
  N.~Itzhaki,
  %``Black hole information versus locality,''
  Phys.\ Rev.\ D {\bf 54}, 1557 (1996)
  doi:10.1103/PhysRevD.54.1557
  [hep-th/9510212].
  
  

%\cite{Chen:2014bva}
\bibitem{Chen:2014bva} 
  P.~Chen, Y.~C.~Ong and D.~h.~Yeom,
  %``Generalized Uncertainty Principle: Implications for Black Hole Complementarity,''
  JHEP {\bf 1412}, 021 (2014)
  %doi:10.1007/JHEP12(2014)021
  [arXiv:1408.3763 [hep-th]].


%\cite{Gim:2017rmn}
\bibitem{Gim:2017rmn} 
  Y.~Gim, H.~Um and W.~Kim,
  %``Black hole complementarity with the generalized uncertainty principle in gravity's rainbow,''
  JCAP {\bf 1802}, no. 02, 060 (2018)
  %doi:10.1088/1475-7516/2018/02/060
  [arXiv:1712.04444 [gr-qc]].



%\cite{Townsend:1997ku}
\bibitem{Townsend:1997ku} 
  P.~K.~Townsend,
  ``Black holes: Lecture notes'',
  gr-qc/9707012.


\bibitem{BB}
S.~J.~Blundell and K.~M.~Blundell,
``Concepts in Thermal Physics",
Oxford, UK: Univ. Pr. (2006) 464 p



%\cite{Vagenas:2018zoz}
\bibitem{Vagenas:2018zoz} 
  E.~C.~Vagenas, S.~M.~Alsaleh and A.~Farag,
  %``GUP parameter and black hole temperature,''
  EPL {\bf 120}, no. 4, 40001 (2017)
 % doi:10.1209/0295-5075/120/40001
  [arXiv:1801.03670 [hep-th]].

%%%%%%%%%%%%%%%%%%%%%%%%%%%%%%%%%%%%%%%%%
%
%
%
%
%
%
%
%
%
%
%
%
%%%%%%%%%%%%
\end{thebibliography}
\end{document}